\documentclass[aps,prd,twocolumn,nopacs,superscriptaddress,nofootinbib,groupedaddress]{revtex4} 

\usepackage{color}
\usepackage{graphicx}
\usepackage{amsmath}
\usepackage{amssymb}
\usepackage[colorlinks=true,citecolor=darkred,urlcolor=darkred, pdfborder={0 0 0}]{hyperref}
\usepackage[normalem]{ulem}
\definecolor{darkred}{rgb}{0.6,0,0}

\definecolor{linkcolor}{rgb}{0,0,0.5}

\def\gsim{\raise0.3ex\hbox{$\;>$\kern-0.75em\raise-1.1ex\hbox{$\sim\;$}}}
\def\lsim{\raise0.3ex\hbox{$\;<$\kern-0.75em\raise-1.1ex\hbox{$\sim\;$}}}

\def\beqn#1{\begin{equation}\label{#1}}
\def\eeqn{\end{equation}}

\def\beqa#1{\begin{eqnarray}\label{#1}}
\def\eeqa{\end{eqnarray}}

\def\C{$\mathcal{C}$}
\def\Z2{$\mathcal{Z_2}$}

\def\hc{\mathrm{h.c.}}
\def\vev#1{\left\langle #1\right\rangle}

\newcommand {\ignore}[1]{}

\newcommand{\AddrAHEP}{%
  AHEP Group, Institut de F\'{i}sica Corpuscular --
  CSIC/Universitat de Val\`{e}ncia, Parc Cient\'ific de Paterna.\\
 C/ Catedr\'atico Jos\'e Beltr\'an, 2 E-46980 Paterna (Valencia) - SPAIN}

\begin{document}

\title{Spontaneous proton decay and the origin of Peccei-Quinn symmetry}
\author{Mario Reig}
\email{mario.reig@ific.uv.es}
\affiliation{\AddrAHEP}

\author{Rahul Srivastava}
\email{rahulsri@ific.uv.es}
\affiliation{\AddrAHEP}



\newcommand {\black} {\color{black}}
\newcommand {\blue} {\color{blue}}
\newcommand {\cyan} {\color{cyan}}
\newcommand {\green} {\color{green}}
\newcommand {\yellow} {\color{yellow}}
\newcommand {\magenta} {\color{magenta}}
\newcommand {\red} {\color{red}}



\begin{abstract}
We propose a new interpretation of Peccei-Quinn symmetry within the Standard Model, identifying it with the axial $B + L$ symmetry i.e. $U(1)_{PQ} \equiv U(1)_{\gamma_5(B+L)}$. This new interpretation retains all the attractive features of Peccei-Quinn solution to strong CP problem but in addition also leads to several other new and interesting consequences. Owing  to the identification $U(1)_{PQ} \equiv U(1)_{\gamma_5(B+L)}$ the axion also behaves like Majoron inducing small seesaw masses for neutrinos after spontaneous symmetry breaking. Another novel feature of this identification is the phenomenon of spontaneous (and also chiral) proton decay with its decay rate associated with the axion decay constant. Low energy processes which can be used to test this interpretation are pointed out.
\end{abstract}

\maketitle

\section{Introduction and motivation}
\label{motivation}

Ever since  it was first proposed in \cite{Peccei:1977hh}, Peccei-Quinn (PQ) symmetry has been the leading and most popular paradigm to explain the so called ``strong CP problem'' i.e. the observed CP conservation by the strong interaction. Its main prediction is the existence of the associated pseudo-Nambu-Goldstone boson, the axion \cite{Wilczek:1977pj, Weinberg:1977ma}. The axion receives its mass from non-perturbative QCD effects (see \cite{diCortona:2015ldu} for a recent review of QCD axion properties) and constitutes one of the most promising dark matter candidates \cite{Preskill:1982cy,Dine:1982ah,Abbott:1982af}. Many experiments are looking for this elusive particle with several present and future experiments in various phases of implementation \cite{Arvanitaki:2014dfa,Barbieri:2016vwg,Stern:2016bbw,Armengaud:2014gea,Majorovits:2017ppy,Giannotti:2017hny,Flambaum:2018wbu,Budker:2013hfa,Irastorza:2018dyq}. 

Despite its great success in addressing the strong CP problem, the origin of PQ symmetry remains a mystery.
In the early days, attempts were made to identify PQ symmetry with the axial baryon number $U(1)_A$.
This implied that the $U(1)_{PQ}$ was spontaneously broken by the Higgs vacuum expectation value (vev) $\vev{\Phi}=v/\sqrt{2}$. The coupling of the axion to matter is, then, inversely proportional to the vev i.e. $f_A^{-1} \propto 1/v$ \cite{Srednicki:1985xd} and was soon ruled out experimentally. The axion can be made \textit{invisible} by assuming that PQ symmetry is broken by the large vev of a Standard Model (SM) singlet scalar. Depending whether the SM quarks are charged under PQ symmetry or not, there exist two main classes of models namely the DFSZ \cite{Dine:1981rt,Zhitnitsky:1980tq} and KSVZ \cite{Kim:1979if,Shifman:1979if} models, respectively. In the DFSZ class of models no additional fermions are required while in the KSVZ models exotic quarks are needed to induce the QCD anomaly for PQ symmetry. However, in either type of invisible axion scenario the important point to note is that the PQ symmetry cannot be associated to axial baryon number like in the original proposals. 

In most of its current avatars the PQ symmetry is assumed to be an additional global symmetry beyond the symmetries present in SM. This symmetry as well as the transformation of the SM fermions under it are fixed in a somewhat ad-hoc fashion. This allows one, for instance, to define models with very different axion-photon coupling \cite{DiLuzio:2016sbl,DiLuzio:2017pfr}. In addition, recently, a number of authors have made some insights in relating the PQ symmetry and the axion to flavor physics  (\cite{Ema:2016ops,Calibbi:2016hwq,Celis:2014jua,Linster:2018avp,Appelquist:2006xd,Arias-Aragon:2017eww,Reig:2018ocz,Berezhiani:1989fp,Berezhiani:1990wn,Bjorkeroth:2018dzu,Bjorkeroth:2017tsz,Alanne:2018fns,Suematsu:2018hbu})\footnote{The idea that PQ symmetry might have a non-trivial relation with a flavor group was proposed a long time ago \cite{Wilczek:1982rv}.}, unification \cite{Ernst:2018bib,DiLuzio:2018gqe,Altarelli:2013aqa,Babu:2015bna,Bajc:2005zf}, inflation (\cite{Freese:1990rb,Banks:2003sx,Pajer:2013fsa, Ballesteros:2016euj,Boucenna:2017fna,Salvio:2015cja} and neutrino mass generation (\cite{Ma:2017vdv,Ma:2017zyb,Ma:2018fhd,Reig:2018ocz,Caputo:2018zky,Huang:2018cwo}).

In this letter we propose a new interpretation of Peccei-Quinn symmetry within SM.  
 It is well known that apart from gauge symmetries the SM also has two accidental symmetries, namely the Baryon $U(1)_B$ and Lepton $U(1)_L$ number symmetries. 
Although both these symmetries suffer from $SU(2)_L$ anomaly, 
it should be noted that neither them nor any linear combination like $U(1)_{B-L}$ or $U(1)_{B+L}$ symmetry can be identified with PQ symmetry. This is simply because they are not anomalous with respect to $SU(3)_c$ symmetry, as needed for the PQ mechanism to work. 

If one relaxes the condition that all SM fermions get mass from only one Higgs doublet, then one can indeed have more general variant of the $U(1)_{B+L}$ symmetry. Of particular interest is the possibility of an axial variant of $U(1)_{B+L}$ symmetry. Here, we
 show that this symmetry, which we denote as $U(1)_{\gamma_5(B+L)}$ can be identified with the PQ symmetry i.e.
\begin{eqnarray}\label{UPQ}
 U(1)_{PQ} \equiv U(1)_{\gamma_5 (B+L)}
 \label{pq=b+l}
\end{eqnarray}
Since the $B$ and $L$ charges of all SM fermions are fixed hence with this identification the PQ charges for different fields can no longer be arbitrary and are automatically fixed to be
\begin{equation}
[q_L] \, = \, -[q_R] \, = \, 1/3\,, \, \, [l_L] \, = \, -[l_R] \, = \, 1 \,.
\label{sm-pq-charges}
\end{equation}
Furthermore, with aim to generate naturally small seesaw masses for neutrinos, we also add three copies of right handed neutrinos with $[\nu_L] = -[\nu_R] = 1$ under the PQ symmetry.

It can be seen that the above PQ charges for quarks coincide with the charges of canonical DFSZ models \cite{Dine:1981rt} up to an unphysical rescaling. Thus the QCD anomaly structure is identical to the DFSZ class of models. Akin to them, the spontaneous breaking of anomalous PQ symmetry drives the QCD vacuum to a CP conserving minimum, thus solving the strong CP problem. The major difference resides in the lepton sector, which in our proposal has a PQ charge three times bigger than quarks. This can potentially allow one to discriminate between axion models and test our proposal.

Just like in usual PQ models, the $U(1)_{\gamma_5(B+L)}$ will be broken spontaneously at high energies. However, the fact that the PQ symmetry is same as axial $B+L$ symmetry leads to several interesting consequences as we discuss in next section.

\section{$U(1)_{\gamma_5(B+L)}$ breaking: The seesaw mechanism and spontaneous proton decay} 

As in usual PQ models, the $U(1)_{\gamma_5(B+L)}$ symmetry has to be broken at a high energy scale. In order to accomplish this we introduce a SM singlet scalar field $\sigma\sim (1,1,0)$ with PQ charge
\begin{equation}
[\sigma] = 2 \,.
\label{sigma-pq-charge}
\end{equation}
The vev of $\sigma$ field will break  $U(1)_{\gamma_5(B+L)}$ to a residual $Z_6$ subgroup under which quarks will 
transform as $\omega$ or $\omega^{-1}$ and leptons will transform as $\omega^3$; $\omega$ being sixth root of unity with $\omega^6 = 1$. 
As we now show, the $\gamma_5(B+L)$ nature of PQ symmetry along with the particular choice of its breaking has the attractive feature that the PQ breaking scale can now be related with the seesaw scale as well as the scale of ``spontaneous proton decay''. To show this we start with the operators responsible for generating Majorana mass of neutrinos and those mediating the proton decay. The Lagrangian responsible for neutrino mass generation is given by:
\begin{equation}
\mathcal{L}_\nu = y^\nu_{ij}\bar{l}^iH^\nu\nu_R^j + \frac{y^M_{ij}}{2}\nu_R^i\nu_R^j\sigma + \hc
\label{maj-mass}
\end{equation}
where $l^i$; $i=1,2,3$ are the SM lepton doublets and $H^\nu$ is a $SU(2)_L$ doublet scalar carrying $\gamma_5(B+L)$ charge of $2$.
The vev of $\sigma$ field apart from breaking the PQ symmetry also simultaneously generates a large Majorana mass $M^{\nu_R} = y^M_{ij} \vev{\sigma}$ for $\nu^i_R$. After EW symmetry breaking, this leads to naturally small masses for SM neutrinos through the well-known type-I seesaw mechanism \cite{Minkowski:1977sc,Yanagida:1979as,Schechter:1980gr}. Notice that this construct is very similar to the canonical Majoron type-I seesaw  models \cite{Chikashige:1980ui,Schechter:1981cv}. The difference being that, in our case it is the spontaneous breaking of $U(1)_{\gamma_5(B+L)}$ which leads to neutrino masses. Thus the resulting axion of our model also behaves like the Majoron of the canonical models. In addition, the connection of PQ symmetry to axial $B+L$ implies a conceptual relation between the axion and neutrino mass scale in a similar way to \cite{Reig:2018ocz}
	
	\begin{equation}
	m_a\sim \frac{\Lambda_{QCD} m_\pi}{v_{EW}^2}m_\nu .
	\end{equation}

Another novel and striking feature is the link between proton stability and PQ symmetry, originating from the fact that in our case PQ symmetry is just $\gamma_5(B+L)$ symmetry. 

Thus the spontaneous breaking of PQ symmetry through $\vev{\sigma}$ also triggers spontaneous proton decay. Indeed this link can be seen at the effective operator level itself. Given the PQ charges of the fermions and $\sigma$, one can see that following effective operators are allowed by all symmetries
\begin{equation}\label{eff_ops}
\frac{1}{\Lambda^3}q_Lq_Lq_Ll_L\sigma^*\,, \quad \frac{1}{\Lambda^3}q_Rq_Rq_Rl_R\sigma\,.
\end{equation}
Once the PQ symmetry is broken by $\vev{\sigma}$, these operators can lead to spontaneous proton decay in an appropriate Ultra-Violet (UV) completion of the model as we will discuss in next section. If we take $\vev{\sigma}\sim 10^{11}$ GeV i.e. the typical PQ breaking scale then one can easily see that not only it concides with a natural seesaw scale but also such a scale implies the UV completion scale to be at or below Grand Unified Theory (GUT) scale. Indeed taking the latest Super-Kamiokande (SK) bounds \cite{Miura:2016krn} we find that
\begin{equation}
\frac{\vev{\sigma}}{\Lambda^3}\leq 4 \times 10^{-32}\,\,\text{GeV}^{-2}\,.
\end{equation}
For $\vev{\sigma}\equiv f_A\sim 10^{11}$ GeV we get 
\begin{equation}
\Lambda\geq 1.36 \times 10^{14}\,\,\text{GeV}\,.
\label{lamb-limit}
\end{equation}
The fact that axial (B+L) is broken in two units is  suggestive of  a particular type of proton decay namely ``chiral spontaneous proton decay''. As we discuss shortly, this opens a new experimental probe to potentially test our proposal in baryon decay experiments. In next section, as an illustrative example, we discuss one of the many possible UV completions of the effective operators in \eqref{eff_ops}.

\section{UV completion}

The effective operators of \eqref{eff_ops} can be easily obtained from a full UV complete model. Since, in our case the PQ symmetry is identified with $\gamma_5(B+L)$ symmetry, any viable UV complete model in its PQ unbroken phase should preserve $U(1)_{\gamma_5(B+L)}$ at the Lagrangian level and to all orders in perturbation theory. It should only be violated by instanton effects, generating the axion potential that drives the theory to a CP conserving minimum.

\begin{figure*}[t]
	\centering
	\includegraphics[width=0.97\textwidth]{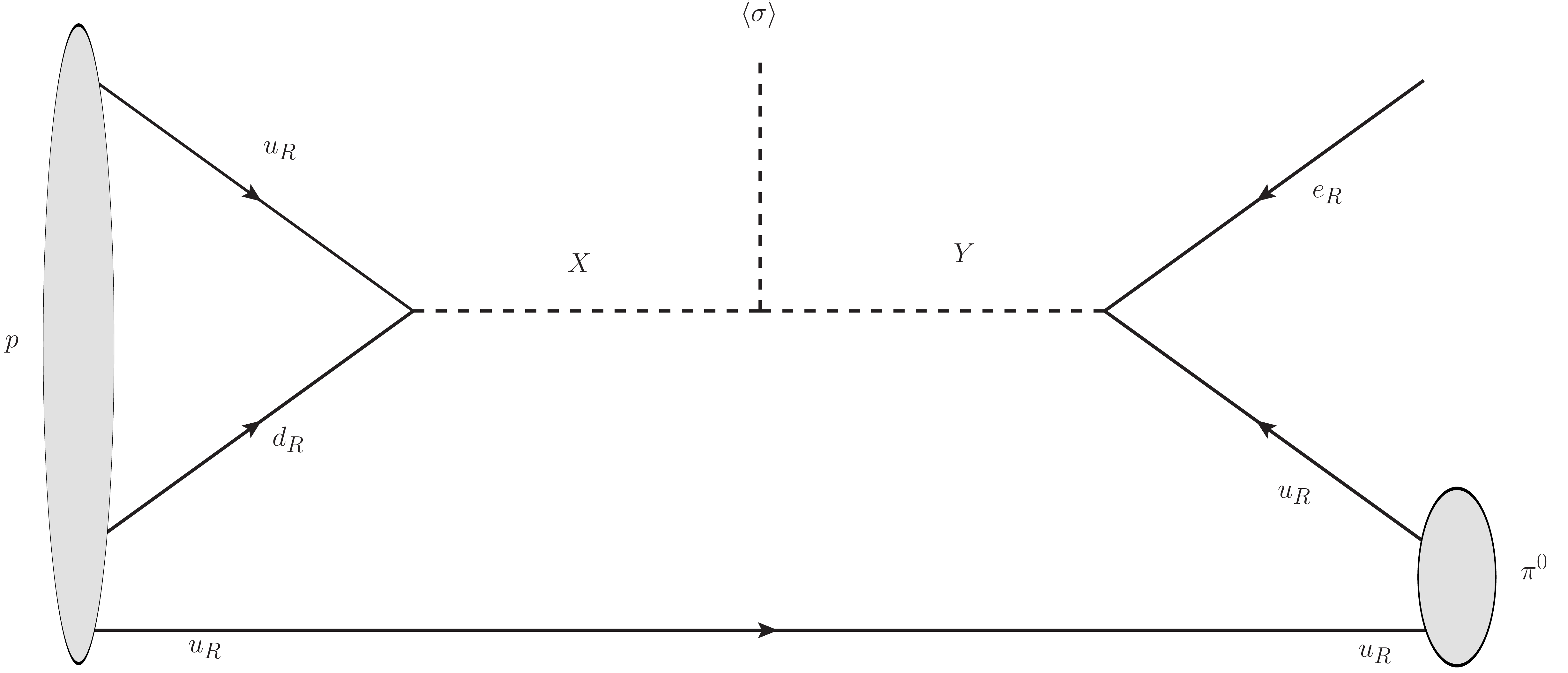}
	\caption{Diagram of spontaneous proton decay.}
	\label{p_decay_diagram}
\end{figure*}

Keeping these requirements in mind, one can UV complete the model in several different ways. For sake of illustration here we discuss only one of the simplest such UV model. Apart from the SM fields, the scalar $\sigma$ and the $SU(2)_L$ doublet scalars required to give all fermions masses\footnote{As we will discuss in coming section, we need four such scalars.}, we introduce a couple of scalar leptoquarks with masses larger than the limit obtained in \eqref{lamb-limit} 
\begin{equation}
X = (3,1,-1/3)_{2/3}\,, \quad Y=(\bar{3},1,1/3)_{4/3}\,,
\label{lepquarks}
\end{equation}
In \eqref{lepquarks} the $SU(3)_c \otimes SU(2)_L \otimes U(1)_Y$ charges of leptoquarks are indicated in parenthesis while the $\gamma_5(B+L)$ charge are in subscripts. The important couplings of these leptoquarks are
\begin{equation}
y_X \, \bar{u}^c_R d_R X \, , \quad y_Y \, \bar{u}^c_R e_R Y \, , \quad  \kappa \, XY\sigma^*\,.
\label{lep-tri}
\end{equation}
where $y_X, y_Y, \kappa$ are coupling constants. Below the mass scale of $X,Y$ they can be integrated out leading to the effective operator $\frac{1}{\Lambda^3} \bar{u}^c_R \bar{u}^c_R d_R e_R \sigma$.
The presence of the couplings in \eqref{lep-tri} implies that the vev of $\sigma$ field not only breaks PQ symmetry but also simultaneously induces
spontaneous as well as chiral (in this case only right handed) proton decay as shown in Fig.\ref{p_decay_diagram}.
%
%
As we remarked before, this particular UV complete model is just one of the many possibilities. Other possible UV complete models can lead to the other operator of 
\eqref{eff_ops} while still other UV completions can result in both operators of \eqref{eff_ops} being present simultaneously. In this short letter we will not go in such details. 
The spontaneous proton decay can in principle be distinguished from other types of proton decay e.g. GUT mediated ones 
in proton decay experiments as we discuss in next section.

\section{Discriminating spontaneous and GUT mediated proton decays}

There are four $d=6$ effective operators leading to proton decay \cite{Wilczek:1979hc}:
\begin{eqnarray}
\mathcal{O}_1 & = & (u_Rd_R)(q_Ll_L)\,, \quad  \mathcal{O}_2 \, = \, (q_Lq_L)(u_Re_R)\,, \nonumber \\
\mathcal{O}_3 & = & (q_Lq_L)(q_Ll_L)\,, \quad  \mathcal{O}_4 \, = \, (u_Rd_R)(u_Re_R)\,.
\end{eqnarray}
In usual GUT proton decay is typically gauge boson mediated. This makes it vectorial and dominated by the operators $\mathcal{O}_1$ and $\mathcal{O}_2$. In contrast, owing to its origin from $U(1)_{\gamma_5(B+L)}$ symmetry, spontaneous  proton decay is chiral, associated exclusively to either purely left-handed or purely right-handed fields. 

Thus, in principle, if proton decay is ever detected, one may be able to distinguish between our model and the usual GUT proton decay. To do that, we define the ratio
\begin{equation}
R=\frac{p\xrightarrow{L}e^+\pi-p\xrightarrow{R}e^+\pi}{p\xrightarrow{L}e^+\pi+p\xrightarrow{R}e^+\pi}\,.
\end{equation}
In the above equation, the $\xrightarrow{R}$ and $\xrightarrow{L}$ refer to the chirality of the final state positrons; right-handed and left-handed, respectively. 
The ratio $R$ is a measure of the number of chiral left handed vs right handed positrons in the final state.

For usual $SO(10)$ GUT gauge mediated proton decay this ratio should be approximately zero \footnote{ In other GUT models R can have calculable non-zero values which depend on the Dynkin indices of the GUT representations in which SM fermions reside.}. For our case it is close to +1 or -1 depending on whether operator $\mathcal{O}_3$ or $\mathcal{O}_4$ dominates. One can easily envisage a general situation where this ratio can also be between the interval $[-1,+1]$ by introducing, in a UV complete model, the appropriate leptoquarks for both left and right handed fields, separately. However, since the corresponding leptoquark Yukawa coupling to matter are in general different for both chiralities, barring fine tuned cancellation, the ratio is expected to be non-zero.

Determination of $R$ requires the discrimination of the helicities of the final state positrons. In current experiments like super-K and the future hyper-K,
helicity measurement is probably difficult.  However, if proton decay is ever observed, then the positron helicity determination can potentially be done by  measuring the circular polarization of its bremsstrahlung \cite{Macq:1958dz}. 

\section{Other Potential Experimental Tests}

We now briefly discuss some other possible experimental tests which can be used to test our proposal.

\subsubsection{Discriminating between axion models}

As we mentioned before, our framework has some similarities with the standard DFSZ model. It has, however, some important differences as we discussed in previous sections. Another major difference that can help to discriminate between our model and the standard DFSZ is the value of the ratio of axion couplings to photons and gluons $E/N$. The DFSZ-like models predict a ratio \cite{Srednicki:1985xd}
\begin{equation}
\left(\frac{E}{N} \right)_{DFSZ}=\frac{8}{3}\,,
\end{equation}
where $E$ and $N$ are the electromagnetic and QCD anomalies respectively. In our case this ratio is predicted to be \footnote{Note that this value can coincide with the DFSZ-III, where there are three $SU(2)_L$ scalar doublets \cite{DiLuzio:2017pfr}. However, our framework is considerably different from it and other tests like spontaneous proton decay can still be used to distinguish our model from DFSZ-III.} : 
\begin{eqnarray}
\left(\frac{E}{N} \right)_{\gamma_5(B+L)} & = & \frac{14}{3} \, .
\end{eqnarray}

The effective coupling induced between the physical axion and photons is given by \cite{diCortona:2015ldu}\begin{equation}
C_{a\gamma}=\frac{\alpha_{EM}}{2\pi f_A}\Big[\frac{E}{N}-1.92(4)\Big]\,,
	\end{equation}
	In our case it is predicted to be 
	\begin{equation}
	C_{a\gamma}^{\gamma_5(B+L)}\approx 3.66 \, C_{a\gamma}^{DFSZ}\,.
	\end{equation}

\subsubsection{Exotic proton and neutrinoless double beta decays}

As mentioned before, due to the $\gamma_5(B+L)$ nature of the PQ symmetry, the pseudo-Nambu-Goldstone boson ($a$) of the model behaves as an axion as well as a Majoron. Thus, one can expect to also have exotic neutrinoless double beta decay involving $a$ emission \cite{Berezhiani:1992cd} given by
\begin{eqnarray}
(A,Z) \to (A,Z+2) + 2e^- + a
\end{eqnarray}

Moreover, since the PQ breaking also triggers spontaneous proton decay therefore one can also expect exotic proton decay again involving $a$ emission given by 
\begin{equation} \label{exotic decay}
p\rightarrow \pi^0\, e^+\,a\,,
\end{equation}
However, for the typical high energy breaking scale of PQ symmetry ($\sim 10^{11}$ GeV) both these processes will be considerably suppressed.

\section{Other Salient features}

Finally we discuss some other salient features of our proposal. 

\subsubsection{Are there extra Goldstone bosons? }

As we remarked earlier, in order to give all SM fermions masses, we require at least four different $SU(2)_L$ doublet scalars with charge assignments given by 
\begin{eqnarray}
H^u & \sim & (1,2,-1/2)_{2/3}\,, \quad H^d \, \sim \, (1,2,1/2)_{2/3}\,, \nonumber \\
H^\nu & \sim & (1,2,-1/2)_{2}\,, \quad H^e \, \sim \, (1,2,1/2)_{2}\,.
\end{eqnarray}
With all these doublets, the potential can exhibit a $U(1)^5$ global symmetry and can have several Nambu-Goldstone bosons. Only two of them, that will correspond to $U(1)_Y\times U(1)_{PQ}$, are desirable. However, these unwanted $U(1)$ symmetries are not present in our case owing to the presence of the following quartic terms 
\begin{eqnarray}\label{quartic}
H^\nu H^e\sigma^\dagger\sigma^\dagger &,& \quad H^uH^dH^dH^{e\dagger}\,, \nonumber\\
H^dH^uH^uH^{\nu\dagger}\,&,&
H^{d\dagger}H^uH^eH^{\nu\dagger}\,.
\end{eqnarray}
These quartic terms break the unwanted $U(1)$ symmetries thus eliminating all other Nambu-Goldstone bosons except the axion.
Thus, we conclude that the only $U(1)$ symmetries at the potential are the hypercharge $U(1)_Y$ and the PQ symmetry $U(1)_{PQ}$ leading to axions.

\subsubsection{Residual discrete symmetry}

As we mentioned before, the vev of the scalar $\sigma$ breaks the continuous $U(1)_{\gamma_5(B+L)}$ symmetry down to a discrete $Z_n$ subgroup. The residual $Z_n$ subgroup can be computed by looking at the PQ charges of the fermions and $\sigma$. Since $\sigma$ has charge +2 and the quarks have charge $1/3$, the unbroken symmetry turns out to be a $Z_6$ with $q_L \sim \omega$ and $q_R \sim \omega^5$ under it; $\omega$ being sixth root of unit with $\omega^6 =1$. All the leptons will transform as $\omega^3$ under the residual $Z_6$. Since $Z_6$ is an even $Z_n$ group and neutrinos transform exactly as $\omega^{n/2} \equiv \omega^3$ under it, therefore they can be Majorana in nature \cite{Hirsch:2017col}. Indeed as we showed earlier, Majorana mass do appear for right-handed neutrinos after breaking of PQ symmetry, in turn inducing the seesaw mass for left handed neutrinos as well. 

Since the scalars $H^u, H^d$ required for generating quark masses also carry PQ charge, therefore after electroweak symmetry breaking they will further break the $Z_6$ down to a $Z_2$ symmetry. The breaking can potentially introduce the well-known domain wall problem \cite{Sikivie:1982qv}. However, just like in DFSZ model it can be solved by standard mechanisms. 

\subsubsection{Absence of FCNC }

Due to presence of several scalars, potentially dangerous flavor changing neutral currents (FCNC) mediated by the scalars can arise. Our framework, however, is free of such dangerous interactions. The reason is that, thanks to the chiral nature of PQ symmetry, each scalar couples selectively to only one kind of fermion. This automatically implies that the mass matrix of each fermion is proportional to its Yukawa coupling matrix
\begin{equation}
M_f\propto Y_f\rightarrow [M_f,Y_f]=0\,.
\end{equation}
Since they commute they can be simultaneously diagonalized, avoiding FCNC.  We like to point out that the absence of FCNC is actually a generic feature of PQ models and is not really unique to our case. However, it is reassuring to see that it is applicable to our case as well.

\section{Axial $(B-L)$ as PQ Symmetry}
One may easily envisage that in a similar way, PQ symmetry can be associated to $\gamma_5(B-L)$ symmetry. In such a case the PQ charges of the leptons and the scalars have to be changed appropriately. While this assignment also works to solve the strong CP problem leading to axions and linking it to neutrino mass generation, it does not offer the possibility to test it through the spontaneous chiral proton decay. In addition the axion-photon coupling will change according to new PQ charge assignments, i.e. since lepton PQ charge flips its sign the anomaly coefficient $E$ changes affecting the ratio $E/N$, now given by 
\begin{eqnarray}
\left( \frac{E}{N} \right)_{\gamma_5(B-L)} & = & -\frac{4}{3} \, . 
\end{eqnarray}
Finally we point that due to presence of the Yukawa couplings and scalar quartic terms like (\ref{quartic}), both symmetries, $\gamma_5(B-L)$ and $\gamma_5(B+L)$, cannot hold at the same time.

\section{Axial baryon number as PQ Symmetry}

One can straightforwardly extend our arguments to associate PQ to the axial version of baryon number, as in the pioneer work of \cite{Peccei:1977hh,Wilczek:1977pj,Weinberg:1977ma}. However, a SM singlet is needed to break $U(1)_{\gamma_5B}$ at high energies. The fact that leptons carry no axial baryon charge renders our model different from the minimal DFSZ, needing an extra SU(2) doublet Higgs $H^{lep}\sim (1,2,1/2)_0$ with no PQ charge, giving mass to the leptons. In this case neutrino mass and axion mass scales are disconnected and the predicted coupling to photons is different, since $E/N=5/3$ in this scenario. Additionally, two body proton decays being  $\Delta(B+L) = 2$ processes, cannot be connected with PQ symmetry breaking.

\vspace{.25cm}
\section{CONCLUSIONS AND OUTLOOK}

We have shown that one can associate the origin of Peccei-Quinn symmetry to the axial version of the well known $U(1)_{B+L}$ symmetry of the Standard Model. Unlike other axion models where PQ symmetry is a new ad hoc symmetry and the charges of SM fermions under it are not fully fixed, the assumption $U(1)_{PQ}=U(1)_{\gamma_5(B+L)}$ makes the model highly predictive. In fact, all the couplings to matter are fixed once the axion decay constant $f_A$ is known. The
conceptual and phenomenological implications of this identification are diverse and intriguing. The $(B+L)$ nature of PQ symmetry implies that the scale of PQ breaking is associated with the seesaw scale of neutrino mass generation. It also implies that the spontaneous breaking of PQ symmetry can also trigger the novel phenomenon of spontaneous chiral proton decay, allowing us to distinguish between our model and other theories such as usual GUTs. 

We also discussed differences with other standard axion models like DFSZ axion model. Thus, if the axion and/or proton decay is ever observed, it will not be a hard task to fully test our proposal.

\vspace{-0.72cm}
\begin{acknowledgments}
We would like to thank Jos\'e W.F. Valle and Frank Wilczek for useful comments. This work is supported by  the Spanish grants SEV-2014-0398 and FPA2017-85216-P (AEI/FEDER, UE), PROMETEO/2018/165 (Generalitat Valenciana) and the Spanish Red Consolider MultiDark FPA2017-90566-REDC. M.R. is also supported by FPU grant FPU16/01907. 

\textit{R.S. will like to dedicate this paper to his fianc\'ee Rati Sharma. }
\end{acknowledgments}

\providecommand{\href}[2]{#2}\begingroup\raggedright\endgroup

\end{document}